\documentclass[aps,twocolumn,showpacs,prl]{revtex4}
\usepackage[dvips]{graphicx}
\usepackage[english]{babel}
\usepackage[T1]{fontenc}
\usepackage{mathrsfs}
\usepackage[tbtags]{amsmath}
\usepackage{amssymb}
\usepackage{amsxtra}
\usepackage{amsopn}
\usepackage{latexsym}
\usepackage[mathcal]{eucal}
\usepackage{mathtools}

\newcommand{\BE}{\begin{equation}}
\newcommand{\EE}{\end{equation}}
\newcommand{\BA}{\begin{align}}
\newcommand{\EA}{\end{align}}

\newcommand{\nn}{\nonumber}

\begin{document}

\title{Perturbative study of Yang-Mills theory in the infrared}

\author{Fabio Siringo}

\affiliation{Dipartimento di Fisica e Astronomia 
dell'Universit\`a di Catania,\\ 
INFN Sezione di Catania,
Via S.Sofia 64, I-95123 Catania, Italy}

\date{\today}

\begin{abstract}

Pure Yang-Mills SU(N) theory is studied in four dimensional space and Landau gauge by a
double perturbative expansion based on a {\it massive} free-particle propagator.
By dimensional regularization, all diverging mass terms cancel exactly in the double 
expansion, without the need to include mass counterterms that would spoil the symmetry 
of the original Lagrangian. The emerging perturbation theory is safe in the infrared and 
shares the same behaviour of the standard perturbation theory in the UV. 
At one-loop, gluon and ghost propagators are found in excellent agreement with the data of lattice
simulations and an infrared-safe running coupling is derived. A natural scale $m=0.6-0.7$ GeV
is extracted from the data for $N=3$.
\end{abstract}


%12.38.Bx       Perturbative calculations
%12.38.Lg	Other nonperturbative calculations (QCD)
%12.38.Aw	General properties of QCD (dynamics, confinement, etc.)
%14.70.Dj	Gluons
%11.15.Tk       Other nonperturbative techniques (gauge field theories)

%12.38.Gc	Lattice QCD calculations (see also 11.15.Ha Lattice gauge theory)
%11.10.Ef       Field Theory: Lagrangian and Hamiltonian approaches
%11.15.-q	Gauge field theories
%12.20.-m       QED
%11.15.Bt       General properties of perturbation theory (gauge theory)
%12.38.-t	Quantum chromodynamics

\maketitle

Our knowledge of QCD is still limited by the lack of a powerful analytical method for the study
of the infrared range.  It is widely believed that perturbation theory breaks down at 
the low-energy scale $\Lambda_{QCD}\approx$ 200 MeV and the study of very important phenomena,
including hadronization and quark confinement, must rely on phenomenological models or
numerical lattice simulations. Even pure Yang-Mills SU(N) theory is still not fully understood
in its infrared limit. Important progresses have been achieved in the last years by developing
new analytical tools\cite{
aguilar10,aguilar8,aguilar9,aguilar13,aguilar14,aguilar14b,binosi15,papa15,papa15b,szcz,
reinhardt04,reinhardt05,reinhardt08,reinhardt11,reinhardt14,sigma,sigma2,gep2,varqed,varqcd,genself,
ptqcd,hoyer,dudal08,sorella15,dudal15,
huber14,huber15g,huber15L,huber15b,huber15c} 
and by simulating larger and larger lattices\cite{bogolubsky,dudal,twoloop}.

A key role is played by the dynamical mass\cite{cornwall} that the gluon seems to acquire in the infrared 
according to almost all non-perturbative studies.
Moreover, it has been shown that the inclusion of a mass by hand leads to a phenomenological
model that can be studied by perturbation theory\cite{tissier10,tissier11,tissier14}.

In a recent paper\cite{ptqcd}, we pointed out that a massive perturbation theory, based on a massive
free-particle propagator can be developed by an unconventional setting of the perturbative method,
without changing the Lagrangian and without adding free parameters that were not
in the original Lagrangian, yielding a first-principle anlytical treatment based on perturbation
theory.

In this paper the idea is developed by a double expansion in dimensional regularization.
Pure Yang-Mills SU(N) theory is studied in four dimensional space and Landau gauge by an 
expansion in powers of the total interaction and of the coupling. By dimensional regularization,
all diverging mass terms cancel exactly in the double expansion, without the need to
include mass counterterms that would spoil the symmetry of the original Lagrangian.
The emerging perturbation theory is safe in the infrared and shares the same behaviour of the
standard perturbation theory in the UV. While the present letter summarizes the main results, more
details of the calculation will be published elsewhere\cite{ympt}.

Let us consider  pure Yang-Mills  $SU(N)$ gauge theory without
external fermions. The Lagrangian can be written as
${\cal L}={\cal L}_{YM}+{\cal L}_{fix}$
where ${\cal L}_{YM}$ is the standard Yang-Mills term and ${\cal L}_{fix}$ is
the gauge fixing term.
Usually the total Lagragian is split into two parts, a free-particle Lagrangian ${\cal L}_0$
that does not depend on the coupling strength $g$,
and an interaction ${\cal L}_{int}$ that contains  ${\cal O}(g)$ and ${\cal O} (g^2)$ terms.
In Landau gauge the gluon free-particle propagator is transverse
\BE
\Delta_0^{\mu\nu}(p)=t^{\mu\nu}(p)\frac{1}{-p^2};\qquad t^{\mu\nu}(p)=g^{\mu\nu}-\frac{p^\mu p^\nu}{p^2}
\EE
and has a pole at $p^2=0$. We can shift the pole and modify the free-particle propagator as
\BE
\Delta_0^{\mu\nu}(p)\to\Delta_m^{\mu\nu}(p)=t^{\mu\nu}(p)\frac{1}{-p^2+m^2}
\label{Deltam}
\EE
without changing the content of the theory provided that the counterterm 
\BE
\delta {\cal L}= \frac{1}{2} m^2 A_{\mu} A^{\mu}
\label{counter}
\EE
is added to the interaction. Actually, we are just adding and subtracting the same quantity in ${\cal L}_0$ and
${\cal L}_{int}$ without changing the total Lagrangian.
Thus we can develop a perturbative expansion in powers of the total interaction and use the standard
formalism of perturbation theory with a total interaction ${\cal L}_{int}+\delta{\cal L}$ that is
a mixture of terms that depend on the coupling strength $g$ and the counterterm that does not vanish
in the limit $g\to 0$. In the expansion, the free-particle propagator is the massive propagator (\ref{Deltam}),
and the vertices are of order ${\cal O} (g^0)$ (the counterterm), order ${\cal O} (g)$ 
(the three-particle gluon and ghost-gluon vertex) and order ${\cal O} (g^2)$ (the four-gluon vertex).
That the content of the theory has not changed can be easily seen by summing up all graphs with $n$ insertions
of the counterterm in a gluon line. As shown in Fig. 1, we can write a dressed propagator as the
infinite sum of a set of reducible graphs
\BE
\Delta(p)=\frac{1}{-p^2+m^2}\sum_{n=0}^{\infty}\left[ m^2 \frac{1}{-p^2+m^2}\right]^n=\frac{1}{-p^2}.
\label{geometric}
\EE
The same result can be described as the effect of a proper polarization term $\Pi=m^2$ that arises 
from the counterterm and cancels the mass. Then formally, the two expansions are equivalent if we sum 
up all graphs.
On the other hand, by the same token, at any finite order, the perturbation theory that we develop is not 
equivalent to the standard perturbation theory, but the two expansions differ by an infinite class of graphs
that introduce non-perturbative effects. In other words, the two expansions differ by some non-perturbative
content at any finite order. In fact, it is well known that the gauge invariance of the theory does not
allow any shift of the pole in the propagator at any finite order, so that the massive zeroth order
propagator $\Delta_m$ cannot be obtained by the standard perturbation theory at any finite order.

Since we already know, by non-perturbative calculations, that the gluon propagator is massive in the infrared,
then we expect that the present expansion, with a massive zeroth-order propagator, would be more
reliable  than the standard expansion, at least in the infrared. 

\begin{figure}[t] \label{fig:counterterm}
\centering
\includegraphics[width=0.09\textwidth,angle=-90]{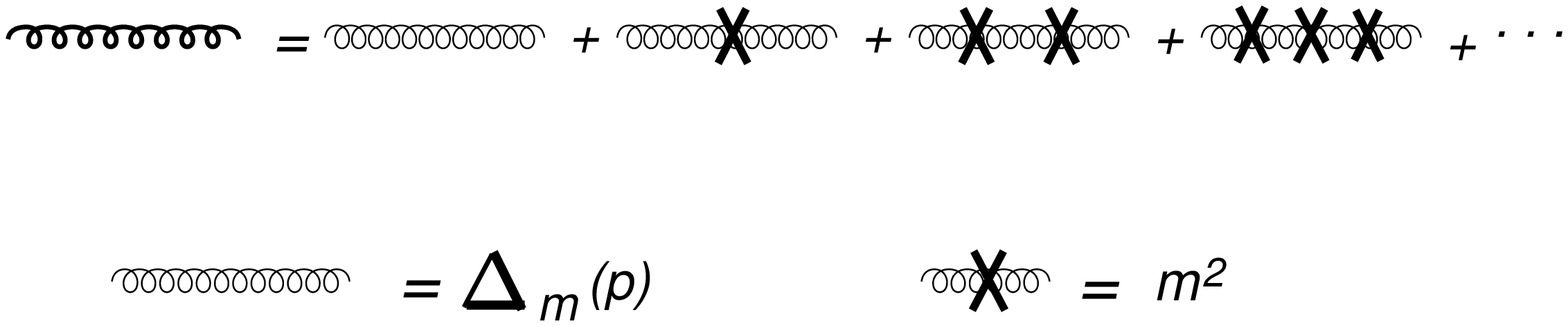}
\caption{Graphical illustration of Eq.(\ref{geometric}). The cross is the counterterm of Eq.(\ref{counter})
that gives a factor $m^2$.}
\end{figure}

While massive models have been studied before 
and found in fair agreement with the lattice data\cite{tissier10,tissier11,tissier14},
the present approach is very different because the Lagrangian is not modified and does not break 
BRST symmetry, so that
no free parameters are added to the exact Yang-Mills theory, 
yielding a description that is based on first principles and can be improved order by order. For instance,
we can easily show that no mass is predicted for the photon by the same method.

Having left the Lagrangian unchanged, we expect that the massive expansion should share the same behaviour
of the standard expansion in the UV where any finite mass becomes negligible. In fact, if $p^2\gg m^2$
the geometric expansion in Eq.(\ref{geometric}) is convergent and the two perturbation theories must
give the same results. On the other hand, when $p^2\to m^2$ each single term of the geometric expansion
Eq.(\ref{geometric}) diverges and the formal sum of infinite poles 
introduces some non-perturbative content\cite{hoyer}.
Thus we can predict that the scale $m$ should be close to the Landau pole $\Lambda$ where the 
standard perturbation theory breaks down.

\begin{figure}[b] \label{fig:graphs}
\centering
\includegraphics[width=0.25\textwidth,angle=-90]{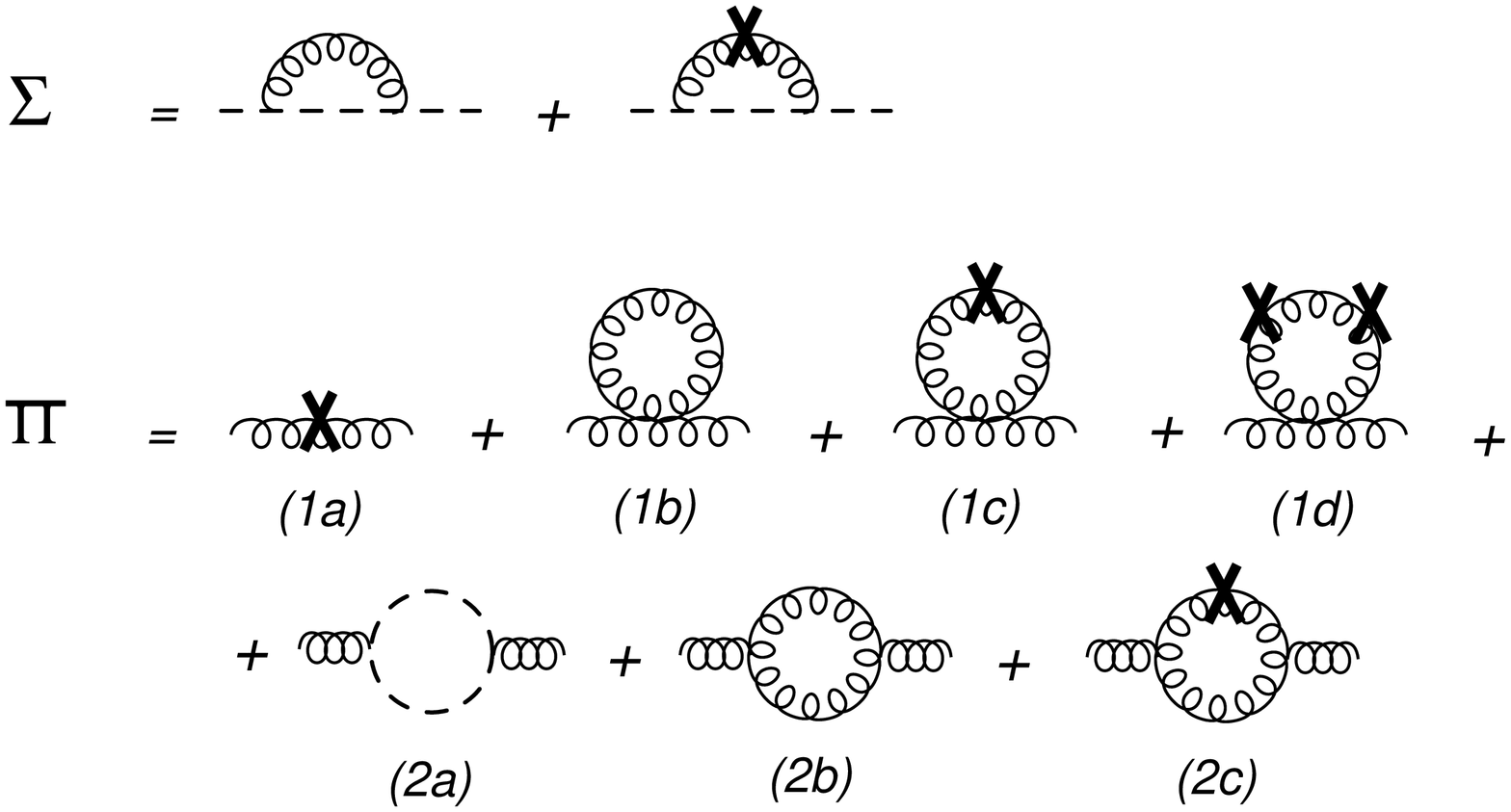}
\caption{Two-point graphs with no more than three vertices and no more than one loop. 
The ghost self energy and the gluon polarization contributing to the functions $F$ and $G$ 
are obtained by the sum of all the graphs in the figure.}
\end{figure}

An other interesting aspect of the present massive expansion is that no other mass counterterms are required. 
Thus there is no need to include terms that, because of gauge invariance, were not in the original Lagrangian. 
All diverging mass terms are cancelled exactly
by the counterterm $\delta {\cal L}$. We  expected that a cancellation like that 
would arise by summing all
graphs, just because the Lagrangian has not been changed and no diverging mass terms are present in the standard
expansion. However, if we inspect the graphs in Fig.2, we can easily see that any insertion of $\delta {\cal L}$
in a loop reduces the degree of divergence of the graph, and all mass terms become finite by a finite number
of insertions. Thus, if the divergences must cancel, they will cancel at a finite order of the expansion
provided that we retain more counterterm insertions than loops.

As pointed out before, the order of a graph is the number of vertices that are included, while
the number of loops is equal to the powers of $g^2$ in the graph. If the effective coupling is small, 
as it turns out to be according to non-perturbative calculations, 
we could consider a double expansion in powers of the total 
interaction and in powers of the coupling: we can expand up to the order $n$, retaining graphs with $n$ vertices
at most, and then neglect all graphs with more than $\ell$ loops. If $n$ is large enough, then all divergences
in the mass terms are cancelled  by the counterterms in the loops. For instance, at one loop 
we only need $n=3$. 

In this paper we report the results for a third-order one-loop double expansion in dimensional regularization.
The gluon polarization and the ghost self-energy are evaluated by the sum of all graphs with no more than
three vertices and no more than one loop, as shown in Fig.2. The integrals are evaluated analytically by
dimensional regularization and expanded in powers of $\epsilon=d-4$. 

In dimensional regularization, the cancellation of all diverging mass-terms can  be easily
proven by a simple argument. The insertion of just one counterterm in a loop can be seen as the
replacement 
\BE
\frac{1}{-p^2+m^2}\to\frac{1}{-p^2+m^2} m^2\frac{1}{-p^2+m^2}=-m^2\frac{\partial}{{\partial}m^2}\Delta_m
\label{deriv}
\EE
in the internal gluon line. If the graph has no other counterterm insertions, then its dependence on
$m^2$ comes only from the massive propagators and a derivative of the whole $n$th-order $\ell$-loop
graph gives the sum of all $(n+1)$-order $\ell$-loop graphs that can be written by a single
insertion in any possible ways. According to Eq.(\ref{deriv}), each diverging mass term $m^2/\epsilon $ 
that comes from a loop would give a crossed-loop term  $-m^2/\epsilon$. The argument also suggests a simple
way to evaluate the crossed-loop graphs by Eq.(\ref{deriv}).

The cancellation of all diverging mass terms without inclusion of any other conterterm is very
important because there is no need to include free parameters that were not in the Lagrangian, while
all other divergences can be dealt with by standard wave function renormalization.

It is instructive to inspect the constant graphs that contribute to the proper gluon polarization $\Pi$
in Fig 2.
At the lowest order ($n=1$, $\ell=0$) the counterterm $\delta {\cal L}$ just adds the 
constant term $\Pi_{1a}=m^2$ that cancels the shift of the pole in the propagator. The tadpole $\Pi_{1b}$
is
\BE
\Pi_{1b}=\frac{3}{4} \alpha\>m^2\>\left(\frac{2}{\epsilon}+\log\frac{\mu^2}{m^2}+\frac{1}{6}\right)
\EE
where the effective coupling $\alpha$ is given by
\BE
\alpha=\frac{3N}{4\pi} \alpha_s;\qquad \alpha_s=\frac{g^2}{4\pi}.
\EE
The crossed tadpole $\Pi_{1c}$ can be evaluated directly or by a derivative according to Eq.(\ref{deriv})
\BE
\Pi_{1c}=-m^2\frac{{\partial}\Pi_{1b}}{{\partial} m^2}=
-\frac{3}{4} \alpha\>m^2\>\left(\frac{2}{\epsilon}+\log\frac{\mu^2}{m^2}-\frac{5}{6}\right).
\EE
The diverging terms already cancel in the sum $\Pi_{1b}+\Pi_{1c}$. In fact, the double-crossed tadpole $\Pi_{1d}$
is finite and including its symmetry factor it reads
\BE
\Pi_{1d}=-\frac{3}{8} \alpha\>m^2
\EE
so that the sum is
\BE
\Pi_{1b}+\Pi_{1c}+\Pi_{1d}=\frac{3}{8} \alpha\>m^2.
\EE
While the ghost loop vanishes in the limit $p\to 0$, a mass term can arise from the gluon loop that
in the same limit is
\BE
\Pi_{2b}(0)=-\alpha m^2\left(\frac{2}{\epsilon}+\log\frac{\mu^2}{m^2}+{\rm const.}\right)
\EE
and adding the crossed loop with its symmetry factor
\BE
\Pi_{2b}(0)+\Pi_{2c}(0)=\left(1-m^2\frac{\partial}{\partial m^2}\right)\Pi_{2b}=-\alpha m^2.
\EE
Thus the one-loop gluon propagator reads
\BE
\Delta(p)^{-1}=-p^2+\frac{5}{8}\alpha m^2-\left[ \Pi(p)-\Pi(0)\right].
\label{Delta}
\EE
We observe that a finite mass term has survived, and it is of order $\alpha$. Since it only arises from the
gluon loops, no mass would survive in QED for the photon by the same method.

The calculation of the total one-loop polarization and of the ghost self energy is straightforward
but tedious. 
For a massive theory, the one-loop sunrise graphs $\Pi_{2a} (p)$, $\Pi_{2b} (p)$ and the one-loop
ghost self energy have been evaluated by several authors. The crossed loops follow by a mass derivative
according to Eq.(\ref{deriv}). In the minimal subtraction scheme, all the divergences are cancelled
by the same wave function renormalization constants of the standard expansion. Namely we obtain
\BE
Z_A-1=\frac{13\alpha}{9\epsilon}; \qquad Z_\omega-1=\frac{\alpha}{2\epsilon}.
\label{Z}
\EE
The resulting propagators are finite, and the ghost and gluon dressing functions, $\chi$ and $J$ respectively,
can be written in units of the scale $m^2$. Without taking any special subtraction point, 
the dressing functions can be recast as
\BE
J(s)=\frac{J(s_1)}{1+\alpha J(s_1)\left[F(s)-F(s_1)\right]}
\label{J}
\EE
\BE
\chi(s)=\frac{\chi(s_2)}{1+\alpha \chi(s_2)\left[G(s)-G(s_2)\right]}
\label{chi}
\EE
where $s=p^2/m^2$. The integration points $s_1$, $s_2$ are arbitrary as
also are the normalization constants $J(s_1)$, $\chi(s_2)$. The functions $F(x)$, $G(x)$
do not depend on any scale or parameter and are given by the following explicit expressions
\begin{align}
F(x)&=\frac{5}{8x}+\frac{1}{72}\left[L_A+L_B+L_C+R_A+R_B+R_C\right]\nn\\
G(x)&=\frac{1}{12}\left[L_G+R_G\right]
\label{dress}
\end{align}
where the logarithmic functions $L_X$ are
\begin{align}
L_A(x)&=\frac{3x^3-34x^2-28x-24}{x}\>\times\nn\\
&\times\sqrt{\frac{4+x}{x}}
\log\left(\frac{\sqrt{4+x}-\sqrt{x}}{\sqrt{4+x}+\sqrt{x}}\right)\nn\\
L_B(x)&=\frac{2(1+x)^2}{x^3}(3x^3-20x^2+11x-2)\log(1+x)\nn\\
L_C(x)&=(2-3x^2)\log(x)\nn\\
L_G(x)&=\frac{(1+x)^2(2x-1)}{x^2}\log(1+x)-2x\log(x)
\label{logs}
\end{align}
and the rational parts $R_X$ are
\begin{align}
R_A(x)&=-\frac{4+x}{x}(x^2-20x+12)\nn\\
R_B(x)&=\frac{2(1+x)^2}{x^2}(x^2-10x+1)\nn\\
R_C(x)&=\frac{2}{x^2}+2-x^2\nn\\
R_G(x)&=\frac{1}{x}+2.
\label{rational}
\end{align}
In the UV the functions $F,G$ have the asymptotic behaviour
\BE
F(x)\approx\frac{17}{18} +\frac{13}{18}\log(x);\qquad G(x)\approx\frac{1}{3} +\frac{1}{4}\log(x)
\label{asympt}
\EE
so that the standard UV behaviour is recovered for $s,s_0\gg 1$
\begin{align}
J(s)^{-1}&=1+\frac{13}{18}\alpha\log(s/s_0)\nn\\
\chi (s)^{-1}&=1+\frac{1}{4}\alpha\log(s/s_0)
\end{align}
as we could predict from the wave function renormalization constants Eq.(\ref{Z}).

In the opposite limit $x\to 0$ we find $G(x)\to {\rm const.}$ and $F(x)\sim (1/x)$ so that $\chi(0)$ is
finite and $J(s)\sim (s/\alpha)$, yielding a finite gluon propagator $\Delta(0)=8/5\alpha m^2$ as predicted
by Eq.(\ref{Delta}).

Up to a renormalization factor, the dressing functions are invariant for a change of the bare coupling.
That is  more evident if we consider the rescaled functions $(\alpha J)$ and $(\alpha \chi)$ since then
Eqs.(\ref{J}),(\ref{chi}) loose any explicit dependence on $\alpha$.
In fact, the physical content of the theory is inside the universal functions $F,G$. 
The actual value of $m^2$ takes the role of a natural scale that fixes the physical units and can only be 
determined by a comparison with
some phisical quantity, as for lattice simulations. That is just a consequence of the lack of an energy
scale in the Lagrangian. However, the present calculation is very predictive and, up to irrelevant constants,
the inverse dressing functions are predicted to take the universal shape of the functions $F$ and $G$: we
can write Eqs.(\ref{J}),(\ref{chi}) as
\begin{align}
\left[\alpha p^2 \Delta(p)\right]^{-1}&=F(p^2/m^2)+{\rm const.}\nn\\
\left[\alpha \chi(p)\right]^{-1}&=G(p^2/m^2)+{\rm const.}
\end{align}
where the constants depend on normalization, bare coupling and subtraction points.
Thus we expect that, up to a scaling factor, all lattice data can be put on top of the plots of $F$ and $G$
by an additive constant. In Fig.3 and Fig.4 the lattice data of Ref.\cite{bogolubsky} are shown together
with the plots of the functions $F$ and $G$. The function $F$ has a pronounced minimum at $x\approx 1.62$ that
is $p\approx0.93$ GeV in the units of the lattice data, thus fixing the scale $m=0.73$ GeV that is used in the
figures.   We also show
the gluon propagator and the ghost dressing function in Fig.5 and Fig.6.

The agreement is very good for a one-loop calculation but deviations can be expected when $s$ is far from
the subtraction point. Thus a slight dependence on the subtraction point would be
a natural consequence of the one-loop approximation.

\begin{figure}[t] \label{fig:F}
\centering
\includegraphics[width=0.35\textwidth,angle=-90]{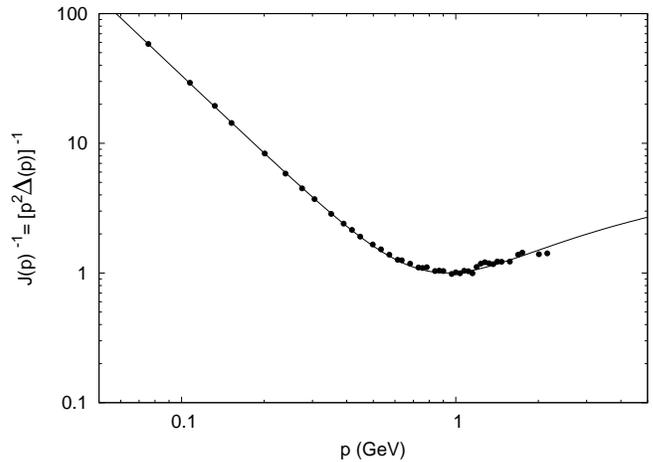}
\caption{The function $F(x)+c$ (line) is plotted together with the
lattice data for the 
inverse gluon dressing function $1/J(p)=p^2\Delta(p)$  (points) extracted from
the figure of Ref.\cite{bogolubsky} ($N=3$, $g=1.02$, L=96) and   scaled
by a renormalization factor. The energy scale is set by taking $m=0.73$ GeV.}
\end{figure}

\begin{figure}[t] \label{fig:G}
\centering
\includegraphics[width=0.35\textwidth,angle=-90]{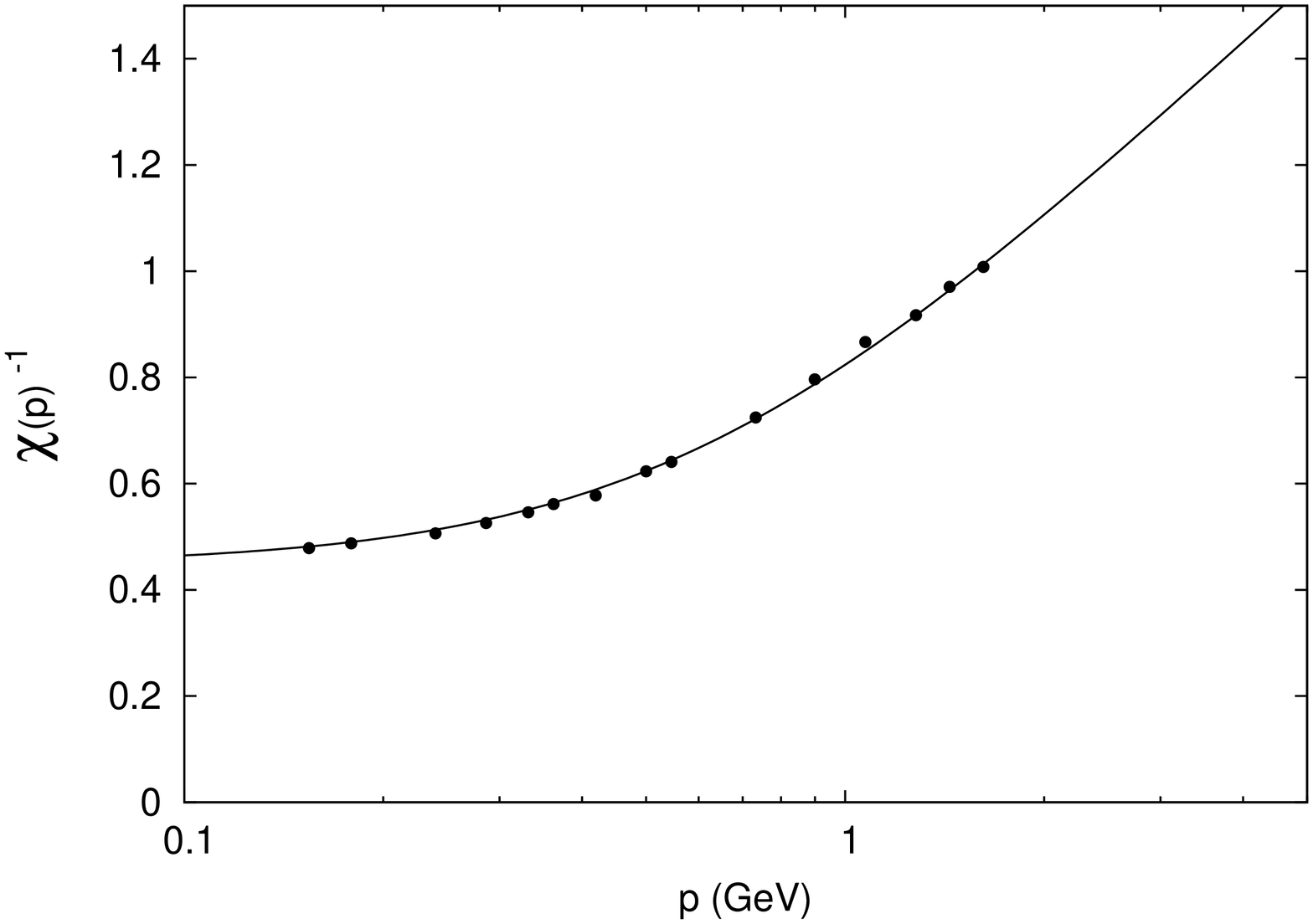}
\caption{The function $G(x)+c$ (line) is plotted together with the
lattice data for the 
inverse ghost dressing function $1/\chi(p)$  (points) extracted from
the figure of Ref.\cite{bogolubsky} ($N=3$, $g=1.02$, L=96) and   scaled
by a renormalization factor. The energy scale is set by taking $m=0.73$ GeV.}
\end{figure}

\begin{figure}[hm] \label{fig:Delta}
\centering
\includegraphics[width=0.35\textwidth,angle=-90]{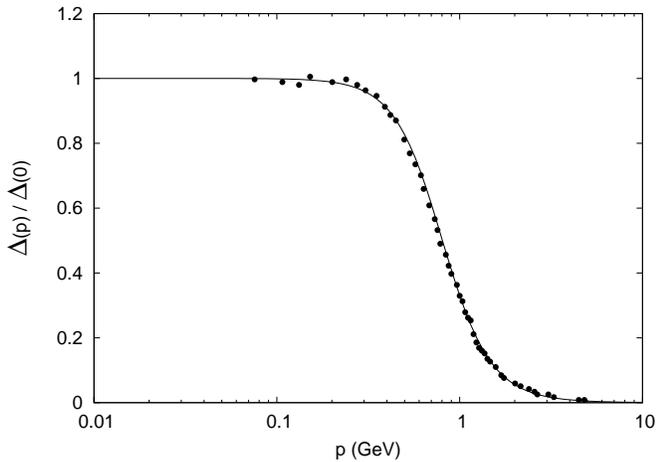}
\caption{The gluon propagator $\Delta (p)$ obtained by Eq.(\ref{J}) (line) 
is plotted together with the lattice data (points) extracted from
the figure of Ref.\cite{bogolubsky} ($N=3$, $g=1.02$, L=96) and   scaled
by a renormalization factor. The energy scale is set by taking $m=0.73$ GeV.}
\end{figure}

\begin{figure}[t] \label{fig:chi}
\centering
\includegraphics[width=0.35\textwidth,angle=-90]{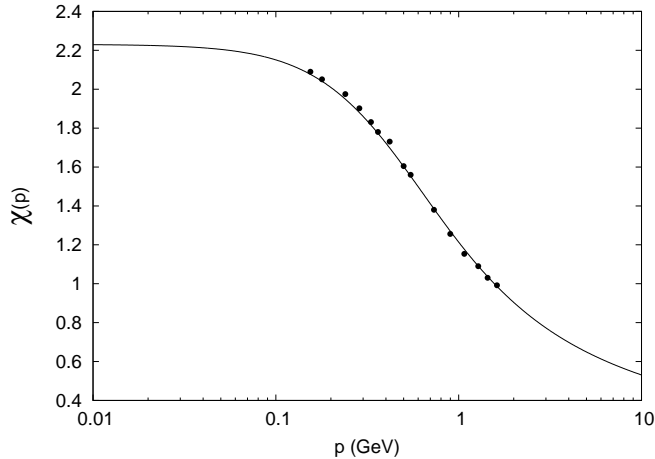}
\caption{The ghost dressing function $\chi (p)$  (line) is plotted together with the
lattice data (points)  extracted from
the figure of Ref.\cite{bogolubsky} ($N=3$, $g=1.02$, L=96) and   scaled
by a renormalization factor. The energy scale is set by taking $m=0.73$ GeV.}
\end{figure}

\begin{figure}[t] \label{fig:alphas}
\centering
\includegraphics[width=0.35\textwidth,angle=-90]{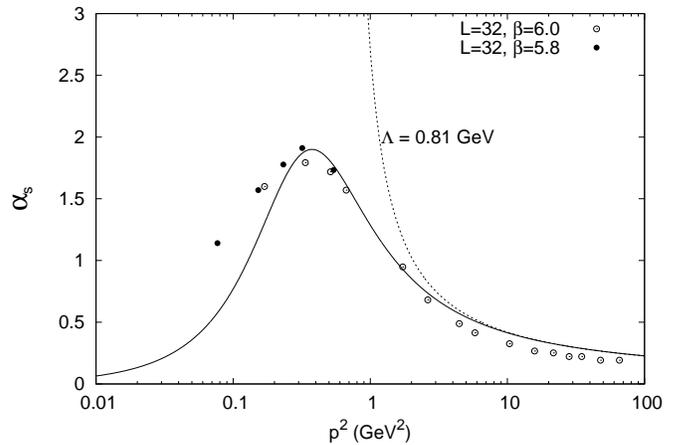}
\caption{The running coupling $\alpha_s=4\pi\alpha/(3N)$ for $N=3$ by Eq.(\ref{alpha}). 
$s_0$ and $\alpha_s(s_0)$ are fixed at the maximum ($s_0=x_M=1.044$, $\alpha_s(s_0)=1.9$) and
$m=0.6$ GeV (solid line). The points are the lattice data of Ref.\cite{twoloop}. 
The broken line is the standard one-loop behaviour for $\Lambda=0.81$ GeV.}
\end{figure}

Nevertheless, let us pretend that we can ignore such limitations and look at observable quantities 
like the running coupling
that can be related to phenomenology. Assuming that in Landau gauge the ghost-gluon vertex 
is regular\cite{taylor} and the vertex renormalization constant can be set to one in a momentum-subtraction
scheme, a runnig coupling is usually defined by the renormalization group invariant
product $\alpha(s)=\alpha_0 J(s) \chi(s)^2$ where $J(s_0)=\chi(s_0)=1$ and $\alpha_0=\alpha(s_0)$.
Using Eqs.(\ref{J}) and (\ref{chi}), the one-loop running coupling can be written as
\BE
\alpha(s)=\frac{\alpha(s_0)}{1+\alpha(s_0)\left[S(s)-S(s_0)\right]}
\label{alpha}
\EE
where $S(x)=F(x)+2G(x)$. 

By Eq.(\ref{asympt}), for $s,s_0\gg 1$ we find the standard UV behaviour
$\alpha[S(s)-S(s_0)]\approx (11\alpha/9)\log (s/s_0)=(11N\alpha_s/12\pi)\log (s/s_0)$. In this limit
the result does not depend on the scale $m$ and we recover the well known one-loop
running coupling. 

In the infrared, as shown in Fig.7, the running coupling is finite, it does not
encounter a Landau pole and vanishes in the limit $s\to 0$ as the power $\alpha(s)\sim s$. 
A maximum is found at the point where ${\rm d}S(x)/{\rm d}x=0$, which occurs at $x_M= 1.044$. 
This point does not depend on any
parameter, and can be used as an alternative method for fixing a physical energy scale if that
energy is {\it measured} somehow. For $N=3$, many lattice simulations predict a maximum at $p\approx 0.6-0.7$
GeV giving a scale $m\approx 0.6$ GeV that is not too far from the value $m=0.73$ that was
extracted from the propagators of Ref.\cite{bogolubsky}. Probably, this small difference would 
be narrowed by the inclusion of higher loops in the calculation. 

While in the UV the asymptotic behaviour of the coupling seems to give no information on the scale $m$,
in the infrared a phenomenological knowledge of the coupling would fix all the free parameters.
Suppose that we pinpoint the value of $\alpha_s$ at its maximum, $\alpha_s \approx 1.9$ according to
Ref.\cite{twoloop}, and let us explore as an exercise
the behaviour of the coupling when we go back towards the UV by Eq.(\ref{alpha}), with all the limitations 
of the one-loop approximation. The maximum at $p\approx 0.6$ gives a scale $m\approx 0.6$ GeV. We can set
$s_0$ at this point $s_0=x_M$ and then take $\alpha_s(s_0)$ as the maximum coupling. 
At the same point $S(x_M)=S_M=3.09$. 
Then, pushing $s$ towards the UV and making use of the asymptotic behaviour Eq.(\ref{asympt}), we can
insert $S(s)\approx 29/18+(11/9)\log(s)$ in Eq.(\ref{alpha}) and write it as a standard one-loop coupling
\BE
\alpha_s (\mu^2 )\approx \frac{12 \pi}{11 N \log\left(\mu^2/\Lambda^2\right)}
\EE
where the scale $\Lambda$ is defined as
\BE
\Lambda = b\> m \>\exp\left[-  \frac{12\pi}{22 N\alpha_s(x_M)}   \right]
\label{LQCD}
\EE
and the coefficient $b$ satisfies $b^2=\exp(9S_M/11-29/22)=3.35$.
Eq.(\ref{LQCD}) provides a direct link between the Landau pole of the
standard one-loop running coupling at $p=\Lambda$ and the infrared parameters, 
namely the mass scale $m$ and
the maximum value of the coupling $\alpha_s(x_M)$.
Inserting the value $\alpha_s(s_0)\approx 1.9$ of Ref.\cite{twoloop} in Eq.(\ref{LQCD})
we obtain $\Lambda=0.81$ GeV for $N=3$, which is not too far from the value $\Lambda=0.7$ GeV that is
used in the same paper for a fit of the lattice data in the UV.
Moreover, we can extract from Eq.(\ref{LQCD}) the pure theoretical bound  $m> \Lambda/ b\approx 0.55\Lambda$. 

Of course, the calculation is just a one-loop approximation. It generalizes the standard one-loop results
to the infrared, but mantains the limitations of neglecting higher order terms.
Moreover, when connecting very different scales, the use of renormalization group techniques becomes
mandatory for a quantitative description and has been shown to be very effective 
in other massive models\cite{tissier10,tissier11,tissier14}.

In summary, the double expansion has been shown to be viable for energies ranging from the UV to
the infrared, without changing the original Lagrangian, reaching a good agreement with the lattice data
from first principles. While the full potentialities of the method are totally unexplored yet,
it might extend the standard perturbative appoach to lower energies deep inside the {\it non-perturbative} 
sector of QCD.

\end{document}